\documentclass[sigplan,twocolumn,review]{acmart}
\renewcommand\footnotetextcopyrightpermission[1]{}
\usepackage[normalem]{ulem}
\usepackage{amsmath}
\usepackage{amsfonts}
\usepackage[utf8]{inputenc}
\usepackage[T1]{fontenc}
\usepackage{tabularx}
\usepackage{listings}
\usepackage{multirow}
\usepackage{setspace}
\usepackage{caption}
\usepackage{array}
\usepackage{ctable}
\usepackage{float}
\usepackage{algorithm}  
\usepackage{algorithmicx}
\usepackage{algpseudocode}  

\usepackage{graphicx}
\usepackage{chngcntr}
\usepackage{newtxtext} 
\usepackage{newtxmath} 
\usepackage{subfig}
\usepackage{balance}

\usepackage{marvosym} 
\usepackage[symbol]{footmisc}

\newcommand{\sys}{FairBatching}

\renewcommand{\textrightarrow}{$\rightarrow$}

\makeatletter
\newcommand{\multiline}[1]{%
  \begin{tabularx}{\dimexpr\linewidth-\ALG@thistlm}[t]{@{}X@{}}
    #1
  \end{tabularx}
}
\makeatother

\lstdefinelanguage{java}{
  morekeywords={abstract,case,catch,class,def,
    do,else,extends,false,final,finally,
    for,if,implicit,import,match,mixin,
    new,null,object,override,package, public,
    private,protected,requires,return,sealed,
    super,this,throw,trait,true,try,
    type,int,double,while,pnew, persistable, atomic, durable\_root},
  otherkeywords={=>,<-,<\%,<:,>:,\#,@},
  sensitive=true,
  morecomment=[l]{//},
  morecomment=[n]{/*}{*/},
  morestring=[b]",
  morestring=[b]',
  morestring=[b]""",
}

\lstnewenvironment{q1}[1][]
{\lstset{
frame=none,
language=java,
aboveskip=3mm,
belowskip=3mm,
showstringspaces=false,
columns=flexible,
breaklines=true,
basicstyle={\footnotesize\ttfamily}, 
tabsize=4,  
numbers=left,
xleftmargin=2em,
framexleftmargin=2em,
#1} }
 {}
 
\begin{document}

\title{FairBatching: Fairness-Aware Batch Formation for LLM Inference}

\author{Hongtao Lyu, Boyue Liu, Mingyu Wu, Haibo Chen}
\affiliation{
\institution{Institute of Parallel and Distributed Systems, Shanghai Jiao Tong University}            
}
 
\newcommand{\TODO}[1]{\textcolor{red}{TODO: #1}}
\newcommand{\var}[1]{\mathit{#1}}
\newcommand{\func}[1]{\textsc{#1}}
\newcommand{\const}[1]{\mathsf{#1}}

\date{}
\begin{abstract}
Large language model (LLM) inference systems face a fundamental tension between minimizing Time-to-First-Token (TTFT) latency for new requests and maintaining a high, 
steady token generation rate (low Time-Per-Output-Token, or TPOT) for ongoing requests. 
Existing stall-free batching schedulers proposed by Sarathi, while effective at preventing decode stalls, introduce significant computational unfairness. 
They prioritize decode tasks excessively, simultaneously leading to underutilized decode slack and unnecessary prefill queuing delays, which collectively degrade the system's overall quality of service (QoS).

This work identifies the root cause of this unfairness: 
the non-monotonic nature of Time-Between-Tokens (TBT) as a scheduling metric and the rigid decode-prioritizing policy that fails to adapt to dynamic workload bursts. 
We therefore propose {\sys}, a novel LLM inference scheduler that enforces fair resource allocation between prefill and decode tasks. 
It features an adaptive batch capacity determination mechanism, which dynamically adjusts the computational budget to improve the GPU utilization without triggering SLO violations. 
Its fair and dynamic batch formation algorithm breaks away from the decode-prioritizing paradigm, allowing computation resources to be reclaimed from bursting decode tasks to serve prefill surges, achieving global fairness. 
Furthermore, {\sys} provides a novel load estimation method, enabling more effective coordination with upper-level schedulers.
Implemented and evaluated on realistic traces, {\sys} significantly reduces TTFT tail latency by up to 2.29$\times$ while robustly maintaining TPOT SLOs, achieving overall 20.0\% improvement in single-node capacity and 54.3\% improvement in cluster-level capacity.

\end{abstract}

\maketitle
\thispagestyle{plain}
\pagestyle{plain}
\sloppy

\section{Introduction}
\label{sec:intro}

Large language model (LLM) inference has become a cornerstone of modern AI services. The inference process is typically divided into two distinct phases: 
a prefill phase, which processes the entire input prompt in parallel to generate an initial KV cache, 
and a sequential decode phase, which auto-regressively generates output tokens.  
To achieve high hardware utilization on GPUs, inference frameworks batch computations from multiple requests. 
Continuous batching~\cite{orca} and chunked prefill~\cite{sarathi} are key techniques that allow new prefill tasks to be merged with ongoing decode tasks, 
forming hybrid batches that improve overall throughput.

However, efficiently scheduling these hybrid batches presents a significant challenge. 
The system must balance the latency requirements of the prefill phase (TTFT) against the token generation throughput demands of the decode phase (TPOT). 
A common strategy, stall-free batching~\cite{sarathi}, aims to balance this by limiting the size of each computational step (via a token budget) and employing a decode-prioritizing policy. 
This ensures decode tasks are included in every batch, preventing generation stalls and bounding the time between tokens (TBT).

Despite its benefits, this work uncovers that stall-free batching introduces unfairness issues in resource allocation between decode and prefill tasks. 
Through empirical analysis, we observe that even under moderate loads, decode tasks often accumulate substantial slack time (exceeding their TPOT requirements), 
while prefill tasks simultaneously experience severe TTFT violations. 
This unfairness stems from its inflexible decode-prioritizing policy:  
during periods with no prefill tasks,
decode tasks monopolize compute resources. 
But when a burst of prefill tasks arrives, its TBT-bound scheduling mechanism still requires a fixed share of resources to decode tasks, starving the prefill ones and leading to violations. 

Furthermore, the practical relevance between strict TBT guarantees and user-perceived latency is diminishing. Modern application frameworks often employ client-side buffering to smooth token delivery, masking intermediate delays~\cite{wang2024agent, ChatGPTNextWeb}. 
Meanwhile, the rise of Chain-of-Thought reasoning~\cite{wei2022chain} and agentic tool-calling~\cite{hou2025model} also reduces the user-perceived impact of individual token delays. 
In contrast, the average output speed, measured by TPOT, remains a crucial and user-facing metric.

Based on the above observations, we design and implement {\sys}, an SLO-aware scheduling framework for LLM inference that aims to achieve fair computation resource allocation. This goal is realized by addressing the following key challenges:

\textbf{Fine-grained SLO Attainment Tracking}: 
Existing systems such as Sarathi rely on TBT, a metric agnostic to historical service progress, thereby overlooking accumulated slack from earlier over-serving. 
To overcome this, {\sys} introduces a novel envelope-line-based SLO tracking mechanism that uniformly captures both TTFT and TPOT requirements in an equivalent form. 
This formulation enables efficient and fine-grained awareness of each request's service progress, facilitating fairer scheduling.

\textbf{Bridging the Gap Between Batch Capacity and Performance Objectives}: 
A critical challenge in designing LLM inference schedulers is the fundamental disconnect between the control variable (batch capacity) and the performance objective (time-based SLOs).
Prior stall-free batching systems like Sarathi rely on static, token-based budgets to manage batch capacity, which serves as a poor and inflexible proxy for actual execution time.
{\sys} bridges this gap with a more accurate time-budget model and dynamically adapts batch capacity by analyzing available slack, enabling higher GPU utilization without SLO violations.

\textbf{Fair and Dynamic Batch Formation}: 
Although {\sys} expresses both prefill and decode tasks under comparable SLO metrics, their inherent differences, such as high concurrency in prefill versus request-internal sequentiality in decode, 
make simple deadline-based greedy packing suboptimal. 
{\sys} introduces a three-phase batch packing strategy that first prioritizes decode tasks at risk of SLO violations, 
then immediately schedules prefill tasks due to their TTFT-critical nature and unpredictable arrival rates, 
and admits non-urgent decode tasks only when spare capacity remains. 
This approach effectively balances fairness and efficiency by explicitly accounting for the distinct characteristics of each request type.

\textbf{Integration with Cluster-Level Schedulers}: 
In production LLM inference systems, centralized cluster schedulers require accurate load estimation for global load balancing, 
yet existing metrics often struggle to simultaneously capture computational cost and SLO state. 
By leveraging its fine-grained request progress tracking and fair scheduling strategy, 
{\sys} exposes a precise node-level load estimate that enables simple yet accurate global load balancing while maintaining SLO adherence across the cluster.

By employing these approaches, {\sys} enhances the resource allocation fairness between prefill and decode tasks, thereby striking a balance between TTFT latency and decode speed. 
{\sys} is implemented on vLLM, a popular LLM inference framework, 
and evaluated on a variety of workloads, showing that it significantly reduces SLO violations and improves overall QoS. Under three production workloads and multiple model configurations, {\sys} reduces TTFT tail latency by up to 2.29$\times$ while preserving TPOT guarantees. With the same SLO requirements, it delivers an average 20.0\% improvement over baselines in the single-node setting. When integrated with a load balancer tailored for {\sys}, {\sys} further boosts cluster-level capacity by 54.3\%.

To summarize, the contribution of this work includes:

\begin{itemize}
    \item A comprehensive analysis to illustrate the unfairness problem in stall-free batching schedulers.  
    (\S\ref{sec:background})
    \item {\sys}, a novel LLM inference scheduler that enforces fair compute allocation between prefill and decode tasks. (\S\ref{sec:design})
    \item Experiments on various latency-critical workloads to show how {\sys} improves applications’ tail latency
    (\S\ref{sec:eval})
\end{itemize}

\section{Characterizing Unfairness in LLM Serving}
\label{sec:background}

\subsection{LLM Inference}

The inference process of large language models (LLMs) mainly contains two phases: \emph{prefill} and \emph{decode}. 
The prefill phase is executed first, and during it, semantic information (in the form of \emph{KV cache}) is generated according to the input prompt. 
Since all the prompt tokens are known in advance, the prefill computation for multiple tokens can be executed in one batch as long as the preceding tokens have been processed.
After the prefill phase, LLMs start the multi-round, self-regressive decode phase. 
During each round, it computes the KV-cache for the last generated token and predicts the next one. 
Due to the dependency constraint between different rounds, decode tasks of one request are typically executed sequentially\footnote{Disregarding speculative decoding~\cite{leviathan2023fast}, which can only potentially improve the concurrency of decode tasks to a limited extent, and does not alter the fundamental sequential nature of decode tasks.}.

Modern GPUs rely on batching multiple computational tasks together to fully utilize the parallel computing capability and amortize the necessary memory access time (e.g., for loading model parameters).
To effectively compute the decode tasks, LLM inference frameworks introduce the concept of \emph{continuous batching}~\cite{orca}, which first prefills the newly arriving request and then batches its decode task with existing ones to achieve high concurrency. 

Modern LLM inference frameworks further employ a technique named \emph{chunked prefill}~\cite{sarathi} to further improve the computational efficiency~\cite{tngtech_blog}. 
It allows prefill tasks to be segmented into smaller chunks and merged into an existing decode batch for execution. 
In addition to merging with decode tasks, prefill tasks are also chunked if they are larger than a preset threshold so that the precompiled CUDA graphs can be reused~\cite{cuda_graphs}. 
Mainstream inference systems like vLLM and SGLang have incorporated the chunked prefill feature~\cite{vllm,zheng2024sglang}. 
\subsection{SLO Metrics for LLM Inference}
\label{subsec:slo-metrics}
Latencies are important metrics for LLM inference. 
Unlike other latency-sensitive scenarios (e.g., serverless applications~\cite{chen2019parties}), merely using the end-to-end time as the overall latency, the autoregressive generation behavior of LLM inference is commonly measured in a finer granularity~\cite{nvidia_metrics,vllm_metrics}. 
This section will introduce three commonly used latency metrics usually used as SLO targets. 

\textbf{\emph{TTFT}.} Time-To-First-Token measures the time from a request's arrival to its first token generation.
From a user's perspective, TTFT is the main contributor to the waiting time from when a user sends a request to when receiving a response, and thus is a crucial metric for user experience~\cite{tngtech_blog}.
In a LLM inference framework, TTFT typically contains two parts: a request's queueing time and prefill time. 

TTFT of request $i$ is defined as: 
\begin{equation*}
\mathit{TTFT}_i = \mathit{OutputTime}_{i,0} - \mathit{ArrivalTime}_i
\end{equation*}
where $\mathit{OutputTime}_{i,j}$ is the generation time of $i$-th request's $j$-th token. 

\textbf{\emph{TBT}.} Time Between Tokens (also known as Inter-Token Latency, ITL) is a per-token metric that measures the time between two consecutive token outputs. 
To maintain a low TBT, the output generation needs to be smooth and continuous. 
Any interruption during generation may lead to a high TBT. 

TBT of request $i$'s $j$-th token is defined as:
\begin{equation*}
\mathit{TBT}_{i,j} = \mathit{OutputTime}_{i,j} - \mathit{OutputTime}_{i,j-1}
\end{equation*}
\textbf{\emph{TPOT}.} Time-Per-Output-Token measures the average time spent on each token output. 
From a user's perspective, TPOT represents the average generation speed. 
If TPOT is larger than a threshold perceivable by a user (e.g., the reading speed), the user experience will decline. 

The TPOT of request $i$'s $j$-th token is defined as:
\begin{equation*}
\mathit{TPOT}_{i,j} = \frac{\mathit{OutputTime}_{i,j} - \mathit{TTFT}_{i}}{j - 1}
\end{equation*}
\subsection{Existing LLM Serving Scheduling}
As mentioned earlier, modern GPUs can improve efficiency by organizing computation in batches, 
requiring the system to decide how to construct the batch per step when handling a series of inference requests. 
When forming a batch, two main issues must be considered: 

\begin{itemize}
\item \textbf{Maximum batch size determination}: the system first calculates the maximum batch size (\emph{max-BS}), which is represented by the maximum number of tokens involved in the batch. 

\item \textbf{Batch composition}: according to the \emph{max-BS} constraint, the system then chooses which tasks (and their corresponding tokens) should be included in the current batch. 
\end{itemize}

Existing LLM inference systems mainly follow two batching strategies. 
Under the \emph{prefill-prioritizing} approach, 
the maximum batch size is configured to be relatively large so that when pending prefill tasks exist, 
the system can complete these prefill tasks within one or a few batches, thereby optimizing Time to First Token (TTFT). 
SGLang's default configuration and early implementation of vLLM\footnote{The recent v1 engine of vLLM has eliminated the distinction between prefill and decode, defaulting to assembling batches in FIFO order. However, under its default configuration, it still maintains a sufficiently large \emph{max-BS}, making prefill prioritization the default behavior. When a prefill request arrives, decode tasks still experience a significant delay.} use this approach. 
Since prefill tasks are compute-intensive and can take up to several seconds, 
a large prefill batch may severely affect the latency of decode tasks.
To address this problem, \emph{decode-prioritizing} systems like Sarathi~\cite{sarathi} propose a method named \emph{stall-free batching} that better balances the prefill throughput and decode latency. 
Sarathi observes that as long as (1) the execution time for each batch remains below the TBT constraint and (2) all active decode tasks are included in every batch, the TBT requirement of decode tasks can be satisfied. 
To this end, it limits the size of each batch so that its estimated execution time does not violate the TBT constraint of any decode tasks. 
Furthermore, during batch formation, decode tasks are prioritized to be included in the batch, and the remaining token budget is allocated to chunked prefill tasks. 
\subsection{Unfairness in LLM Serving}
\begin{figure}[t]
    \centering
    \includegraphics[width=\columnwidth]{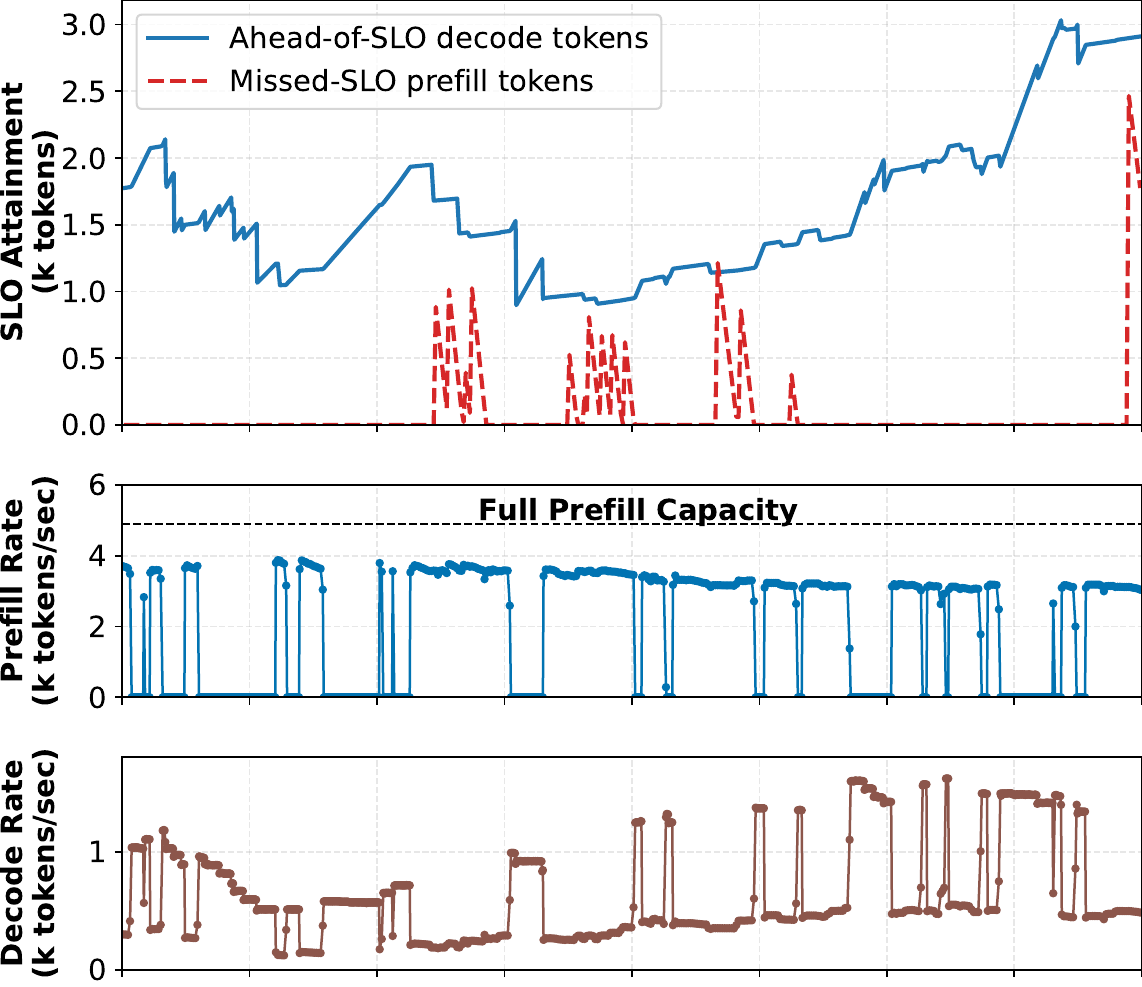}
    \caption{Unfairness in stall-free batching: decode tasks accumulate slack during prefill idleness but still constrain prefill capacity during prefill bursts, causing TTFT violations.}
    \label{fig:moti_relax_vioation_cp}
\end{figure}

Although stall-free batching is currently considered a best practice for balancing decode latency and prefill throughput~\cite{nvidia_tensorrt_llm_chunked_prefill,vllm_chunked_prefill}, 
we have observed unfair computational resource allocation between prefill and decode tasks. 
This unfairness issue leads to decode tasks' latency targets being overly met, while the prefill tasks experience unnecessary queuing delays simultaneously. 
To illustrate this imbalance, 
we implement Sarathi's stall-free batching policy in vLLM and conducted an experiment using the Qwen-Bailian Anonymous Dataset~\cite{kvcache} and the Qwen3-14B model. 
In this experiment, the TTFT and TPOT target is set to 500ms and 50ms, respectively. 
The workload is set to a medium level to observe SLO compliance.
Figure~\ref{fig:moti_relax_vioation_cp} shows the SLO attainment in a token granularity. 
For decode tasks, we find that their token generation rates typically exceed the minimum required by TPOT. 
At any given point in time in Figure~\ref{fig:moti_relax_vioation_cp}, the aggregate progress of all decode tasks is at least 1,000 tokens ahead of the TPOT target. 
However, prefill tasks periodically fall behind their TTFT targets, even when the decode tasks have substantial headroom.

\begin{figure}[h]
    \centering
    \includegraphics[width=\columnwidth]{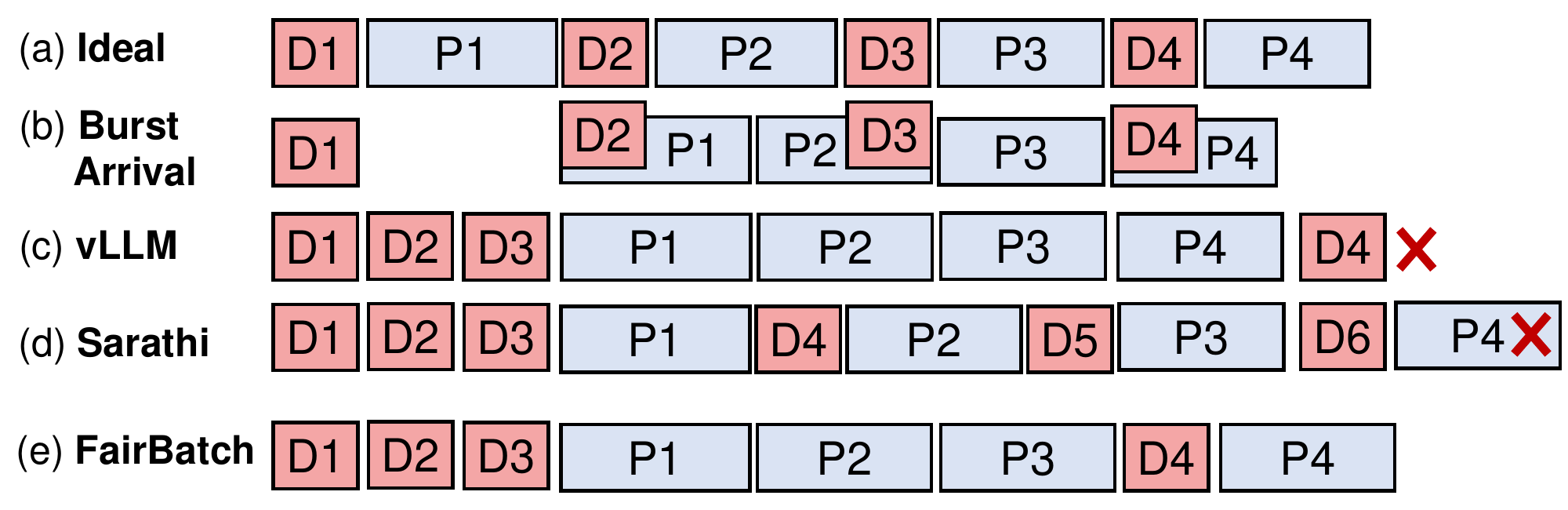}
    \caption{Decisions made by inference systems. (a) Ideal, steady request arrival with fair resource sharing. (b) Realistic bursty arrivals create resource contention periods and idle periods. (c) vLLM's prefill-first policy causes decode starvation. (d) Sarathi's strict TBT guarantees causes TTFT violations. (e) FB allows prefill bursts without breaking TPOT guarantees, achieving the best overall service quality.
    }
    \label{fig:TBT_is_bad}
\end{figure}

Figure~\ref{fig:moti_relax_vioation_cp} further analyzes this unfairness issue by showing the prefill/decode throughput during execution. 
Due to the inherent variability in request arrivals in LLM inference~\cite{wang2025burstgpt}, the system frequently alternates between intervals with no prefill tasks and bursts of prefill tasks. 
Under Sarathi's stall-free batching policy, 
decode tasks monopolize all available compute resources when no prefill tasks exist. 
In contrast, when prefill tasks are present, 
decode tasks are still guaranteed a considerable share of compute resources, hindering prefill tasks from reaching the maximum capacity. 
This phenomenon explains why decode tasks can retain faster progress than their TPOT targets, while prefill tasks fall behind, especially when facing bursts. 

Existing Sarathi-style schedulers struggle to resolve the unfairness issue mainly because they rely solely on TBT as the core metric for decode performance, which is an overly restrictive metric for latency management and is not compatible with fairness. 
First, TBT exhibits a counter-intuitive characteristic: an earlier generation of some tokens in a request can potentially leads to worse TBT attainment. 
As shown in Figure~\ref{fig:TBT_is_bad}(e), although it only generates D2 and D3 earlier compared to the ideal case,
a larger gap is created between D3 and D4, 
resulting in worse TBT attainment. 
To satisfy such a non-monotonicity constraint, Sarathi must maintain the inference progress during prefill bursts, 
even if it has accumulated additional decode progress during prefill idle periods. 
This leads to a continuous accumulation of slack in decode tasks and causes the prefill task to exceed its TTFT. 
Consequently, a violation occurs solely due to uneven request distribution, even though the total computational demand remains unchanged from the ideal scenario (as illustrated in Figure~\ref{fig:TBT_is_bad}(d)). 

However, such TBT violations do not necessarily impact the user experience, as existing LLM frameworks and frontend systems like LangGraph~\cite{wang2024agent} and NextChat~\cite{ChatGPTNextWeb} have integrated smoothing mechanisms. 
These mechanisms automatically cache and reorganize consecutive output tokens, presenting them to the user at a steady rate, thereby masking occasional token-level delays from the upstream inference system.
Furthermore, the growing adoption of Chain-of-Thought (CoT) reasoning~\cite{wei2022chain} and tool-augmented LLM inference~\cite{hou2025model} also reduces the perceived importance of TBT targets. 
In contrast, ensuring a stable TPOT has emerged as a more meaningful SLO for user experience~\cite{tngtech_blog}.

Motivated by the observations above, this work aims to develop an inference system that simultaneously provides TPOT latency guarantees and fair computational resource allocation between prefill and decode tasks.

\section{Design of FairBatching}
\label{sec:design}

This work provides {\sys}, an SLO-aware inference framework to achieve fairness on computation resource allocation between prefill and decode tasks. 
{\sys} receives predefined SLO targets (TTFT and TPOT) as requests' input, propagates them directly into the inference backend, and continuously tracks the SLO compliance status. 
Compared with existing prefill-prioritizing and decode-prioritizing approaches, {\sys} achieves \emph{adaptive batch capacity}, which dynamically determines each batch's capacity according to accurate estimation of per-batch execution time and active requests' SLO targets. 
Meanwhile, {\sys} provides \emph{fair batch formation}, which leverages the slack time accumulated by decode tasks to maximize the efficiency of prefill ones. 
Lastly, we also show that {\sys} can be easily integrated with cluster-level schedulers to improve the performance of distributed LLM inference. 
\subsection{SLO Formulation in {\sys}}

To realize the system described above, the first step is to establish a well-defined SLO metric that can be used for scheduling. 

{\sys} employs an envelope-based approach to define its Service Level Objectives (SLOs). 
For a given SLO target represented by TTFT and TPOT, there exists a set of output time series that satisfy the target. 
These series lie within a certain envelope of curves. 
We adopt the outermost boundary of this envelope as the deadline of each token in a request.  
Specifically, the deadline of request $i$'s $j$-th token is defined as:
\begin{equation*}
    \emph{\text{{token\_ddl}}}_{i,j} = \emph{\text{arrival}}_i + \emph{\text{ttft\_slo}} + \emph{\text{tpot\_slo}} \times j
\end{equation*}

Owing to its independence from current inference progress, 
this per-token SLO deadline exhibits strong monotonicity: 
generating a token earlier will invariably improve the overall SLO compliance rate. 

During scheduling, the scheduler requires each request’s current deadline, which is defined as the target completion time for the next output token. Based on this deadline and the request’s current progress, we can define its \emph{slack time} as follows:
\begin{equation*}
    \emph{\text{{request\_ddl}}}_i = \emph{\text{{token\_ddl}}}_{i,next\_output\_idx}
\end{equation*}
\begin{equation*}
    \emph{\text{{slack}}}_i = \emph{\text{{request\_ddl}}}_i - \emph{\text{{current\_time}}}
\end{equation*}

The slack time quantifies the extent to which a decode task’s progress is ahead of its SLO target, thereby establishing a foundation for a more equitable allocation of computational resources between prefill and decode tasks.

\subsection{Batch Capacity Determination}
\label{subsec:batch-capacity-determination}

The next question to address is how large a batch can be. 
Existing approaches leverage a static token-based mechanism to determine the batch size. 
However, this mechanism fails to adapt to dynamic changes in prefill/decode workloads, leading to the unfairness issue observed in Figure~\ref{fig:moti_relax_vioation_cp}. 
Meanwhile, a batch's execution time cannot be accurately estimated solely by the token budget, making the system fail to allocate adequate resources to prefill and decode tasks. 
To this end, {\sys} \emph{adaptively} determines the capacity of each batch with a \emph{time-based} budget. 

\textbf{Adaptive time-based budget.} 
Unlike existing work determining the number of tokens processed in each batch (i.e., token budgets), {\sys} determines per-batch execution time according to the SLO requirements of all active requests. 
Based on the aforementioned SLO metrics, {\sys} sets a batch's maximum execution time so that it is no larger than the smallest deadline slack of all decode requests. 

However, when facing decode bursts, decode tasks may have very limited deadline slack, and directly using it would cause too small batches and poor performance. 
Therefore, we require the batch size to be larger than the minimum TPOT SLO of active decode requests. 
The final calculation for the time budget is as follows. 
The original token budget is preserved, but it is configured with a larger value solely to ensure compatibility with the CUDA graph's size constraint.

\[
\mathit{init\_time\_budget} = \max\left(\min\limits_i(\mathit{slack}_i), \min\limits_i(\mathit{tpot\_slo}_i)\right)
\]

\textbf{Batch execution time estimation. }
An accurate estimation of per-batch execution time is critical to ensure it falls in the range of the time budget. 
A strawman estimation would still be using the number of new tokens to be processed (token budgets). 
However, our evaluation reveals that this approach can lead to errors of up to $\pm$5.2\%. 
This inaccuracy stems from the fact that while the FFN (Feed-Forward Network) stage of LLM inference does scale computationally with the number of new tokens, 
the attention operators, which are often memory-bandwidth bound~\cite{zhu2025nanoflow}, involve substantial additional access to the KV cache, a cost proportional to the total context length of all participating requests.

To address the inaccuracy problem while maintaining moderate performance overhead on scheduling, {\sys} models the execution time using a linear formulation. 
This model is built offline according to historical performance data and continuously calibrated to ensure accuracy. 
The general form of the model to represent the per-batch time estimation is as follows:
\begin{equation*}
    \emph{\text{{batch\_time}}} = \emph{\text{{a}}} + \emph{\text{{b}}} * \emph{\text{{total\_new\_tokens}}} + \emph{\text{{c}}} * \emph{\text{{total\_context}}}
\end{equation*}
Where \emph{total\_context} and \emph{total\_new\_tokens} model the cost of accessing KV caches and processing tokens, respectively. 
While this modeling approach abstracts away certain low-level GPU execution details, it significantly improves prediction accuracy. 
On the same set of models and traces, the error rate was reduced to $\pm$1.3\%.

\subsection{Fair and Dynamic Batch Formation}
\label{subsec:fair-and-dynamic-token-scheduling}

When the maximum batch capacity for the next step is determined, 
the next decision involves selecting which active tasks should be included in the corresponding batch.
According to the current capacity, this problem can be categorized into three cases:

\textbf{Sufficient Capacity}: 
If all available prefill and decode tokens can fit in the current batch, no special scheduling is required: 
the batch includes all tasks to maximize both inference efficiency and SLO compliance.

\textbf{Moderate Capacity}: 
When the computational demand of all active tasks exceeds the batch capacity, some tasks should be deferred. 
In this case, candidates for deferral are decode tasks with substantial SLO slack or prefill tasks with long remaining TTFT margins. 
{\sys} prioritizes deferring non-urgent decode tasks first, as the set of active decode tasks is relatively stable and predictable, 
whereas delaying prefill tasks, which are subject to uncertain request arrival patterns, poses a higher risk of degrading the system's SLO attainment.

\textbf{Constrained Capacity}: 
If the available capacity is even more scarce, the scheduler must make trade-offs between decode tasks with limited slack time and prefill tasks. 
In such cases, FairBatching prioritizes decode tasks that would otherwise be unable to meet their deadlines if skipped in the current step. 
Specifically, decode tasks whose slack time is smaller than the batch's estimated execution time and the minimum TPOT target of all active tasks are prioritized (\emph{init\_time\_budget + min\_tpot\_slo}), as excluding them is likely to cause a SLO violation in the next step. 
In contrast, delayed prefill tasks can often be handled more flexibly. 
For example, in a distributed inference system with data parallelism, prefill requests can be forwarded to other less-loaded nodes, in the form of tokens, reducing the risk of TTFT violations. 
On the other hand, migrating an active decode task requires either recomputing its KV cache or transferring it between nodes, which is a costly operation that introduces significant overhead. 
Thus, {\sys} conservatively ensures that urgent decode tasks are always included in the current batch. 
This approach also ensures that under extreme workloads, the system can gracefully fall back to a behavior similar to Sarathi's scheduling strategy.

Based on the above analysis, {\sys} implements its batch formation algorithm with a reversed priority order. 
In a preprocessing stage, all under-processing requests are categorized into three groups:
(1) decode tasks with \emph{slack} less than \emph{init\_time\_budget + min\_tpot\_slo},
(2) prefill tasks, and
(3) the remaining decode tasks.
Each group is internally sorted by \emph{slack} in ascending order. 
The scheduler then iteratively selects tasks from the highest to the lowest priority and adds them to the next batch until either the batch capacity is exhausted or all eligible tasks have been included. 
The details of this algorithm are provided in Algorithm~\ref{alg:envelopeserve}.

\noindent \textbf{\emph{\underline{Discussion:}}}
Through this design, {\sys} achieves a more equitable distribution of compute resources between prefill and decode tasks compared to existing systems. 
Although {\sys} retains a certain bias toward decode tasks in principle, 
this bias manifests as full decode-prioritizing behavior only under extreme workloads. 
In normal cases, only decode tasks that are truly urgent are prioritized, 
while the remaining majority of compute capacity is allocated to prefill tasks to absorb unpredictable prefill bursts. 
This design achieves more fair resource allocation between prefill and decode tasks, thereby better aligning with the inherent nature of inference systems to alternate between prefill-burst (when new requests continuously approach) and decode-burst phases (when the prefill queue is idle). 

\begin{algorithm}
\caption{FairBatching Scheduling Algorithm}
\label{alg:envelopeserve}
\begin{algorithmic}[1]
\Procedure{BatchConstruction}{}
    \State \textbf{Input:} 
    \State \quad $\var{active\_requests}$: List of all active requests
    \State \quad $\var{arrival}_i$: Arrival time of request $i$
    \State \quad $\var{ttft\_slo}_i$: TTFT SLO for request $i$
    \State \quad $\var{tpot\_slo}_i$: TPOT SLO for request $i$
    \State \quad $\var{next\_output\_idx}_i$: Next output index for request $i$
    \State \quad $\var{new\_tokens}_i$: Computable new tokens in request $i$
    \State \quad $\var{context}_i$: Context length for request $i$
    \State \quad $\var{model\_params}$: Linear model parameters $a, b, c$

    \State \textbf{Output:} 
    \State \quad $\var{batch} \gets \emptyset$: Selected requests for next step

    \For{\textbf{each} req$_{i} \in \var{active\_requests}$}
        \State $\var{token\_ddl}_{i,j} \gets \var{arrival}_i + \var{ttft\_slo}_i + \var{tpot\_slo}_i \times j$
        \State $\var{request\_ddl}_{i} \gets \var{token\_ddl}_{i,\var{next\_output\_idx}_i}$
        \State $\var{slack}_{i} \gets \var{request\_ddl}_{i} - \var{current\_time}$
    \EndFor

    \State $\var{min\_slack} \gets \min (\var{slack}_i)$
    \State $\var{min\_tpot\_slo} \gets \min (\var{tpot\_slo}_i)$
    \State $\var{init\_time\_budget} \gets \max(\var{min\_slack}, \var{min\_tpot\_slo})$

    \State $\var{group\_ud} \gets \emptyset$ \Comment{Urgent decode tasks}
    \State $\var{group\_p} \gets \emptyset$ \Comment{Prefill tasks}
    \State $\var{group\_nd} \gets \emptyset$ \Comment{Non-urgent decode tasks}

    \For{\textbf{each} $req_i \in \var{active\_requests}$}
        \If{$req_i$.is\_decode \textbf{and} $\var{slack}_i < \var{init\_time\_budget} + \var{min\_tpot\_slo}$}
            \State Add $req_i$ to $\var{group\_ud}$
        \ElsIf{$req_i$.is\_prefill}
            \State Add $req_i$ to $\var{group\_p}$
        \Else
            \State Add $req_i$ to $\var{group\_nd}$
        \EndIf
    \EndFor

    \State Sort $\var{group\_ud}$, $\var{group\_p}$, $\var{group\_nd}$ by $\var{slack}_i$

    \State $\var{time\_budget} \gets \var{init\_time\_budget} - a$  
    \State $\var{token\_budget} \gets {MAX\_TOKEN\_BUDGET}$
    \For{\textbf{each} $req_i \in [\var{group\_ud}, \var{group\_p}, \var{group\_nd}]$}
        \State $\var{time\_cost}_i \gets b \times \var{new\_tokens}_i + c \times \var{context}_i$
        
        \If{$\var{time\_cost}_i \leq \var{time\_budget}$ 
            \textbf{and} $\var{new\_tokens}_i \leq \var{token\_budget}$}
            \State Add $i$ to $\var{batch}$
            \State $\var{time\_budget} \mathbin{\texttt{-=}} \var{time\_cost}_i$
            \State $\var{token\_budget} \mathbin{\texttt{-=}} \var{new\_tokens}_i$
        
        \ElsIf{$\var{token\_budget} > 0$ \textbf{and} $c \times \var{context}_i \leq \var{time\_budget}$}
            \State $\var{cp} \gets \min(\var{token\_budget}, \frac{\var{time\_budget} - c \times \var{context}_i}{b})$
            \State Add $req_i$.chunk($cp$) to $\var{batch}$
            \State $\var{time\_budget} \mathbin{\texttt{-=}} b \times \var{cp} + c \times \var{context}_i$
            \State $\var{token\_budget} \mathbin{\texttt{-=}} \var{cp}$
        \EndIf
    \EndFor

    \State \Return $\var{batch}$
\EndProcedure
\end{algorithmic}
\end{algorithm}

\subsection{Integrated with Upper-Level schedulers}
\label{subsec:upper-level-scheduler-friendly}

The design of {\sys} is also beneficial when applied to a distributed inference system. 
Data parallelism (DP) is a common deployment mechanism for those systems. 
In a DP-based deployment, each DP \emph{rank} corresponds to an independent inference engine with its own context, local lifecycle, and a local scheduler that determines the work to be executed in each step.
Above these inference engines, an upper-level (global) scheduler performs load balancing by routing incoming requests to suitable inference engines, 
preventing any single node from becoming overloaded and causing performance degradation. 

Designing an effective load-balancing algorithm is challenging for two reasons. 
First, since the algorithm must distribute requests for the entire cluster, it cannot be overly complex (especially given that modern LLM inference systems are typically implemented in Python). 
Second, the load balancer must make routing decisions based on the state of each inference instance. 
However, a consistency gap exists between the moment a request is dispatched and when the downstream instance receives it and updates its state. 
This gap is non-trivial: the coordinator and inference engines often reside on different physical machines, 
and even when co-located, multiple engines are typically run as separate processes, complicating state synchronization.

To circumvent the overhead of maintaining strong consistency, a practical approach is to reduce the inference system's state to simple metrics. 
The upper-level scheduler can then update these metrics locally based on previously reported values and its own scheduling decisions, 
ensuring that its routing decisions remain approximately aligned with the system state without incurring large synchronization overhead.
For example, in the latest version of vLLM (v0.10.1.1), its load-balancing algorithm uses request count, defined as a linear combination of waiting and running request counts in each node. 
While this approach overlooks critical details such as the prompt length of each request, the maintenance overhead is trivial and thus can be frequently updated to achieve approximate state consistency with local inference engines. 

The design of {\sys} presents new opportunities to improve the accuracy of the upper-level scheduler's scheduling decisions while maintaining simplicity. 
As {\sys} ensures strong decode latency guarantees at each local engine, the upper-level scheduler can focus primarily on whether a node can accommodate incoming prefill demands within its TTFT SLOs. 
Moreover, {\sys}'s per-batch execution time predictor can be extended to determine how many tokens can be processed within a given time budget. 

\begin{figure}[t]
    \centering
    \includegraphics[width=\columnwidth]{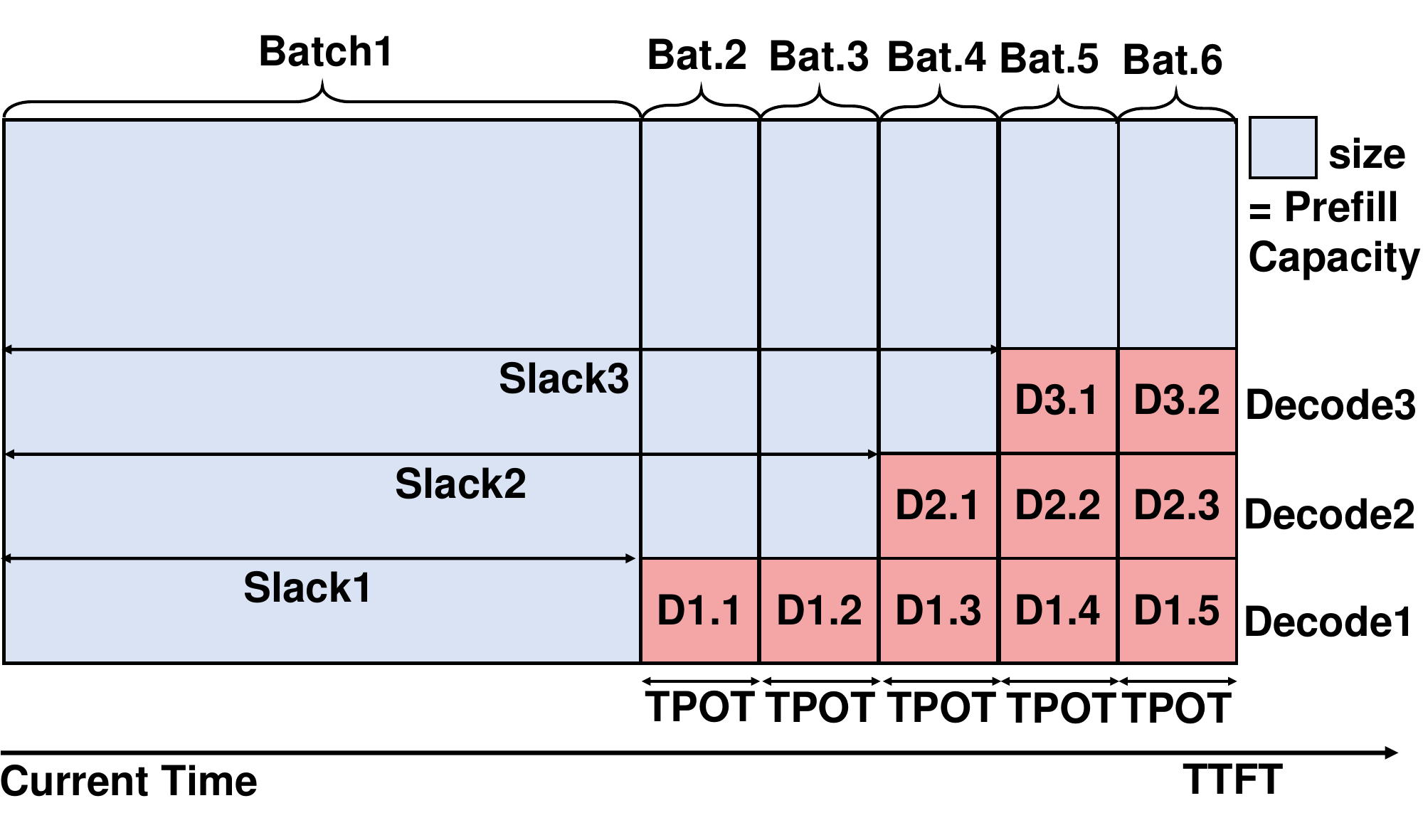}
    \caption{The prefill capacity is calculated as the remaining capacity after subtracting the computational resources already allocated to decode tasks. The prefill admission budget is the prefill capacity minus the already allocated prefill tasks.}
    \label{fig:prefill_capacity}
\end{figure}

To this end, we design a two-level scheduling mechanism for DP-based distributed inference. 
For each batch, each local scheduler with {\sys} calculates the additional number of prefill tokens that can be processed without violating existing SLO (TTFT/TPOT) targets. 
The result is continuously reported to the upper-level scheduler as a prefill admission budget. 
For each incoming request, the upper-level scheduler calculates the number of tokens required for its prefill task and selects a suitable node with sufficient budget for processing. 
After dispatching, the upper-level scheduler decrements the corresponding budget in its local view for subsequent scheduing, and the value will soon be updated in the next batch. 
To calculate the prefill admission budget, {\sys} employs a worst-case relaxation method. 
It estimates a batch's capacity by assuming all decode tasks are delayed until their slack time is exhausted, which maximizes the computational resources reserved for prefill tasks.  
After the prefill task is finished, the local engine switches into a decode burst phase with small batches to avoid TPOT violation. 
As illustrated in Figure~\ref{fig:prefill_capacity}, {\sys} predicts multiple batches that could be executed within the TTFT SLO target required by the incoming request, and all areas not consumed by decode tasks can be used by prefill ones. 
The prefill admission budget is then calculated by subtracting the resources already consumed by active prefill tasks. 
According to {\sys}'s batch time estimation strategy, the prefill admission budget (PAB) is defined as: 

\[
\begin{aligned}
PAB = & \frac{1}{b + c} \left[ TTFT_{slo} - \left[\frac{ (TTFT_{slo} - \min\limits_{i} slack_i)}{TPOT_{slo}}  + 1 \right]  \cdot a \right. \\
& \left. - \sum_{i} \frac{TTFT_{slo} - slack_i}{TPOT_{slo}} \cdot (b + context_i \cdot c) \right] \\
& - \sum_{i \in Prefill} prompt_i
\end{aligned}
\]

Where \emph{$TTFT_{slo}$} and \emph{$TPOT_{slo}$} are global SLO targets. 
A detailed  derivation is provided in Appendix~\ref{appendix:pab}. 

\section{Implementation}
\label{sec:implementation}

{\sys} is implemented based on the popular LLM inference system vLLM v0.10.1.1 (released on August 20, 2025, which was the second-newest version of vLLM at the time of submission; we ported our early version to this version to ensure alignment with the most recent vLLM design). 
{\sys}'s scheduler functions as a V1-compatible scheduler for the vLLM V1 engine (a recent refactored version of vLLM engine) and is developed as an independent, pluggable scheduler module consisting of 1,639 lines of Python code.

To facilitate deployment, {\sys} maintains compatibility with vLLM's existing data structures and execution model. 
The required modifications to the core vLLM framework are thus minimal, 
involving only 90 lines of invasive changes, primarily to propagate per-request SLO information into the scheduler.

At the cluster level, {\sys} extends vLLM's built-in DP-based load balancing (DPLB) framework by replacing its original scheduling metric with the proposed Prefill Admission Budget (PAB). 
Since most of the new logic is encapsulated within {\sys}'s scheduler, integrating this enhancement into the DPLB coordinator required only 16 lines of modification to introduce an alternative scheduling policy.

Additionally, we developed a 2,777-line Python-based automated testing framework. 
This framework is designed to automatically profile vLLM's runtime behavior, perform linear regression, and determine step-time modeling parameters for each model and GPU type.

Another noteworthy implementation detail involves addressing periodic Python GC. 
We observed that, at low frequency, the system would experience pauses on the order of hundreds of milliseconds. 
Analysis revealed that these pauses were caused by the stop-the-world (STW) behavior of Python's garbage collector (GC).
To mitigate GC interference with {\sys}'s control logic, we adopted a two-fold approach. 
First, following existing vLLM practices, we used Python's gc.freeze() function to move long-lived objects into a permanent generation, reducing their impact on GC cycles. 
Second, {\sys} proactively triggers a GC cycle during periods of low load, when there are no queued prefill tasks, decode tasks have ample slack, and a significant time has passed since the last collection. 
This strategy helps avoid overlapping garbage collection with request bursts, thereby minimizing performance disruption.

\section{Evaluation}
\label{sec:eval}

In this section, we present a series of experiments on {\sys}, aiming to answer the following questions:

\emph{\underline{1. Single-Node Performance}}: How much performance improvement does {\sys} achieve when deployed as a single-node scheduler?

\emph{\underline{2. Latency Detail}}: What changes occur in the detailed execution behavior of LLM inference when using {\sys}?

\emph{\underline{3. SLO Versatility}}: How do the performance benefits of {\sys} vary under different SLO targets compared to the baseline systems?

\emph{\underline{4. Performance Breakdown}}: Where do these gains originate from?

\emph{\underline{5. Cluster-Level Performance}}: When integrated into a cluster and using PAB for load balancing, how much does {\sys} improve overall cluster-wide inference efficiency?

\subsection{Experimental setup}
We evaluate {\sys} on a set of popular models with different GPU configurations, using three distinct datasets for testing.
We compare {\sys} against different baselines to show the performance benefits of {\sys}.

\textbf{Model and hardware configurations.} To demonstrate {\sys}'s adaptability across varying model architectures and hardware configurations, we evaluated the system using three distinct models paired with three purpose-built hardware setups, as detailed in Table \ref{tab:model_and_hardware_configurations}. 

\begin{table}[h]
    \caption{Model and hardware configurations}
    \centering
    \begin{tabular}{@{}c  c@{}}
        \hline
        Model & GPU Configurations \\
        \hline
        LLaMA3.1-8B & 1 $\times$ A800-80GB-PCIe \\
        Qwen3-14B & 1 $\times$ H20-96GB-SXM \\
        Qwen3-32B & 2 $\times$ H20-96GB-SXM \\
        LLaMA3.3-70B & 4 $\times$ H20-96GB-SXM \\
        \hline
    \end{tabular}
    \label{tab:model_and_hardware_configurations}
\end{table}

\begin{figure}[t]
    \centering
    \includegraphics[width=1\columnwidth]{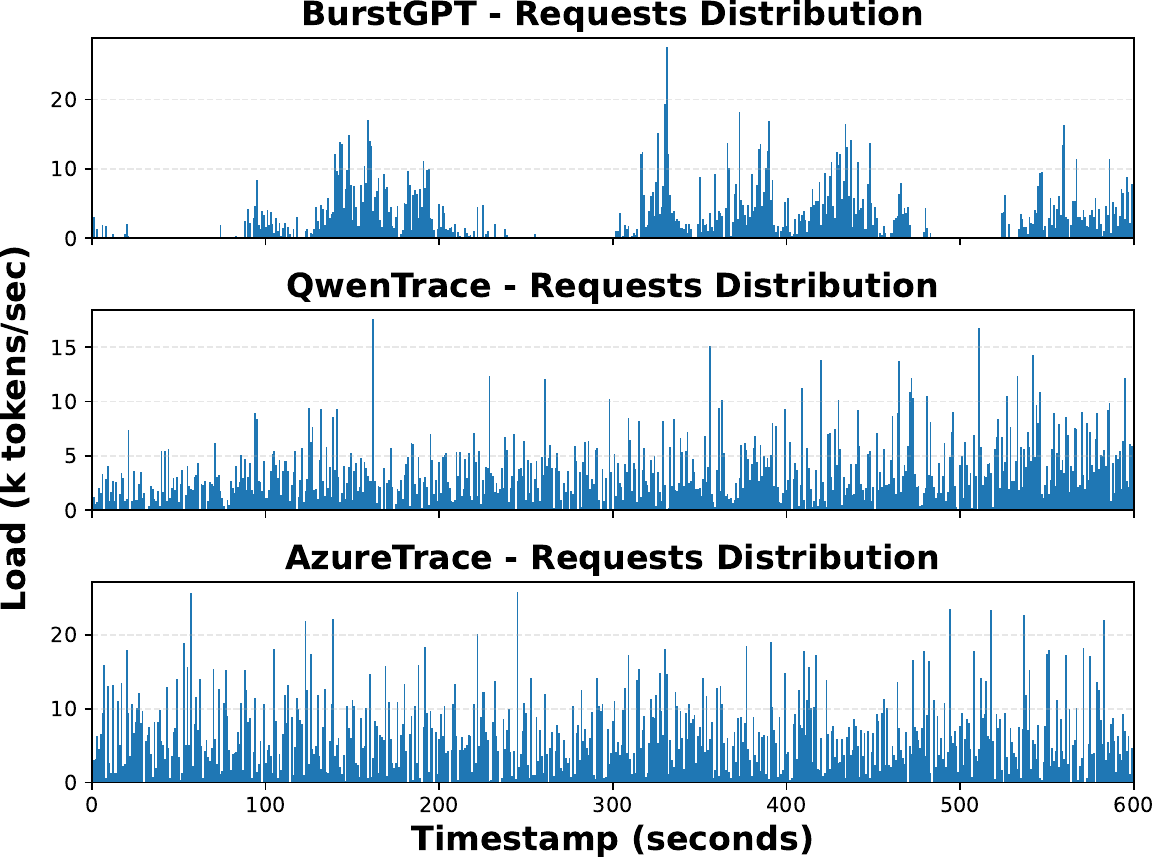}
    \caption{Dataset requests distribution}
    \label{fig:dataset_requests_distribution}
\end{figure}

\textbf{Dataset and SLO.}
To validate {\sys}'s effectiveness under diverse workload patterns, we evaluated the system under three production traces from real enterprise deployments: BurstGPT~\cite{wang2025burstgpt}, Qwen-Bailian Anonymous Dataset~\cite{kvcache}, and Azure LLM inference trace 2024~\cite{stojkovic2025dynamollm}.
By replaying these traces at different scaling factors, we assessed {\sys}'s performance across a range of load intensities.
Given the excessive length of the original traces, we extracted the head segments for testing. 
The request arrival distribution of each truncated trace is summarized using frequency histograms in Figure~\ref{fig:dataset_requests_distribution}. 
The three datasets exhibit distinct input and output length characteristics. 
For instance, due to its longer input sequences, the AzureTrace dataset was assigned a more relaxed TTFT SLO; detailed information is provided in Table~\ref{tab:dataset_configurations}.

\begin{table}[t]
    \caption{Dataset characteristics}
    \centering
    \begin{tabularx}{1\columnwidth}{@{}c *{6}{X}@{}}
        \hline
        \multirow{2}{*}{Dataset} 
        & \multicolumn{2}{c}{Prompt Len.} 
        & \multicolumn{2}{c}{Output Len.} 
        & \multicolumn{2}{c}{SLO (ms)} \\
        \cline{2-3} \cline{4-5} \cline{6-7}
        & avg. & p90 & avg. & p90 & ttft & tpot \\
        \hline
        BurstGPT & 688 & 1599 & 237 & 470 & 500 & 50 \\
        QwenTrace & 892 & 1776 & 377 & 742 & 500 & 50 \\
        AzureTrace & 1604 & 3561 & 114 & 392 & 2000 & 50 \\
        \hline
    \end{tabularx}
    \label{tab:dataset_configurations}
\end{table}

\textbf{Tested systems.}
We compare two versions of {\sys} against different baselines to show the performance benefits of {\sys}.
The tested systems include:
\begin{itemize}
    \item \underline{\emph{vLLM-vanilla}}: vLLM under default configuration, which is a prefill-prioritized serving system.
    \item \underline{\emph{vLLM-sarathi}}: vLLM equipped with a Sarathi-style stall-free batching strategy. This is a decode-prioritized serving system. The most important hyperparameter, TokenBudget, is best tuned for each testcase.
    \item \underline{\emph{{\sys}-vanilla}}: a modified version of vLLM, using {\sys} scheduler, supporting AdaptiveBatchCapacity and DynamicBatchFormation.
    \item \underline{\emph{{\sys}-PAB}}: a variant of {\sys} with additional Prefill Admission Budget (PAB) mechanism for proactive load control. It safeguards against system overload by rejecting new task admissions when prefill capacity is nearing exhaustion. In single-node experiments, {\sys}-PAB is used to emulate the performance when the upper-level scheduler allocates load reasonably to the lower-level instances.
\end{itemize}

\begin{figure*}[!t]
    \centering 
    \subfloat[Llama-3.1-8B; BurstGPT]{%
        \includegraphics[width=0.24\linewidth]{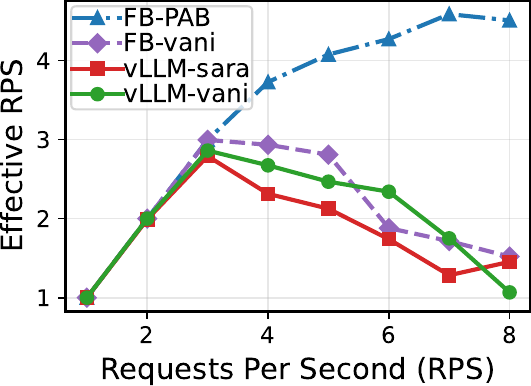}
        \label{fig:subfig_8B-BurstGPT}
    }
    \hfill
    \subfloat[Qwen3-14B; BurstGPT]{%
        \includegraphics[width=0.24\linewidth]{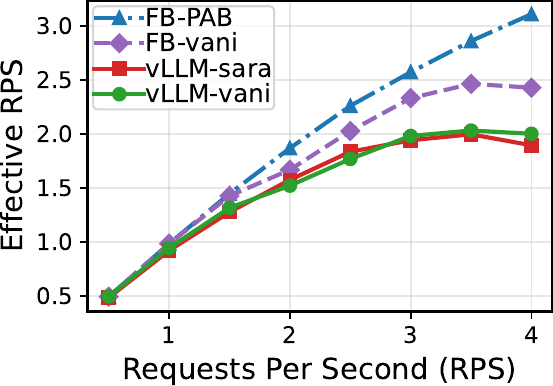}
        \label{fig:subfig_h}
    }
    \hfill
    \subfloat[Qwen3-32B; BurstGPT]{%
        \includegraphics[width=0.24\linewidth]{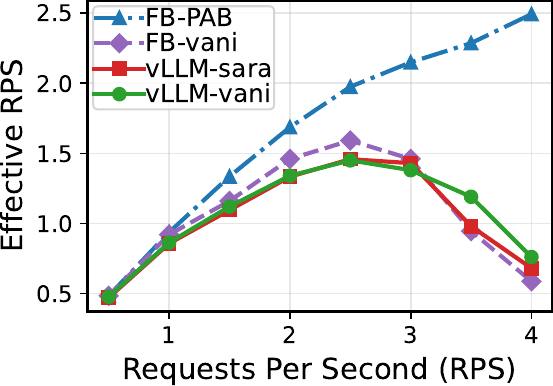}
        \label{fig:subfig_i}
    }
    \hfill
    \subfloat[Llama-3.3-70B; BurstGPT]{%
        \includegraphics[width=0.24\linewidth]{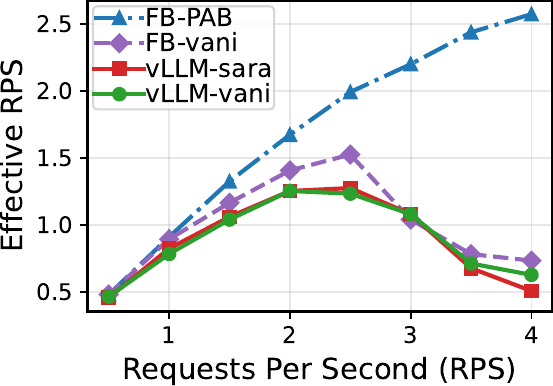}
        \label{fig:subfig_g}
    }\\
    \subfloat[Llama-3.1-8B; QwenTrace]{%
        \includegraphics[width=0.24\linewidth]{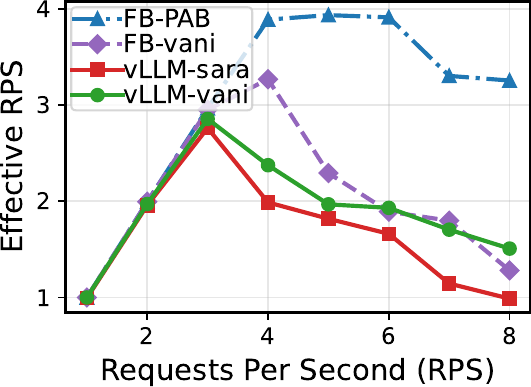}
        \label{fig:subfig_8B-QwenTrace}
    }
    \hfill
    \subfloat[Qwen3-14B; QwenTrace]{%
        \includegraphics[width=0.24\linewidth]{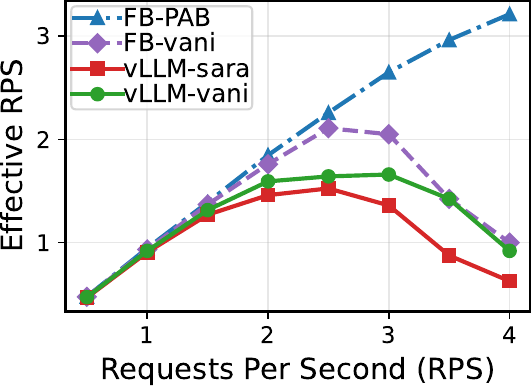}
        \label{fig:subfig_b}
    }
    \hfill
    \subfloat[Qwen3-32B; QwenTrace]{
        \includegraphics[width=0.24\linewidth]{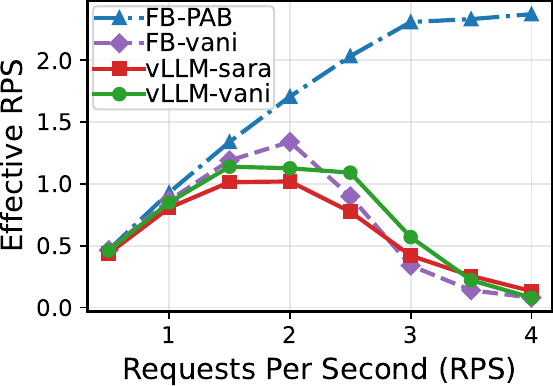}
        \label{fig:subfig_c}
    }
    \hfill 
    \subfloat[Llama-3.3-70B; QwenTrace]{
        \includegraphics[width=0.24\linewidth]{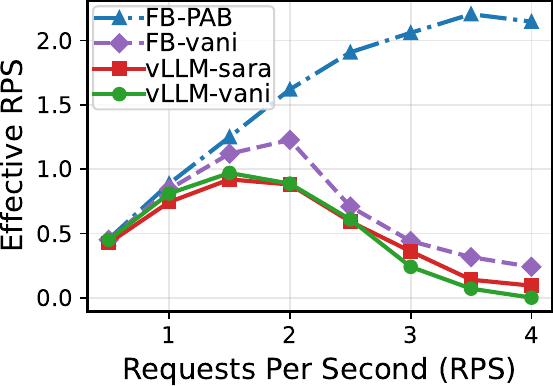}
        \label{fig:subfig_a} 
    }\\
    \subfloat[Llama-3.1-8B; AzureTrace]{
        \includegraphics[width=0.24\linewidth]{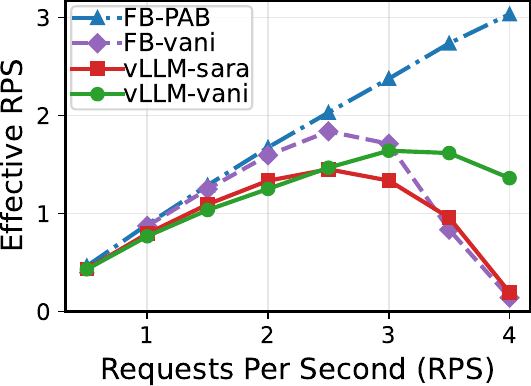}
        \label{fig:subfig_8B-AzureTrace}
    }
    \hfill
    \subfloat[Qwen3-14B; AzureTrace]{
        \includegraphics[width=0.24\linewidth]{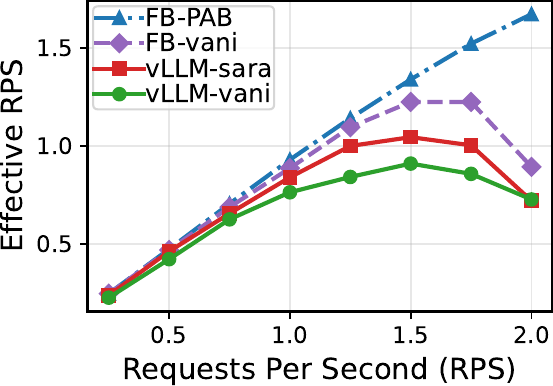}
        \label{fig:subfig_e}
    }
    \hfill
    \subfloat[Qwen3-32B; AzureTrace]{
        \includegraphics[width=0.24\linewidth]{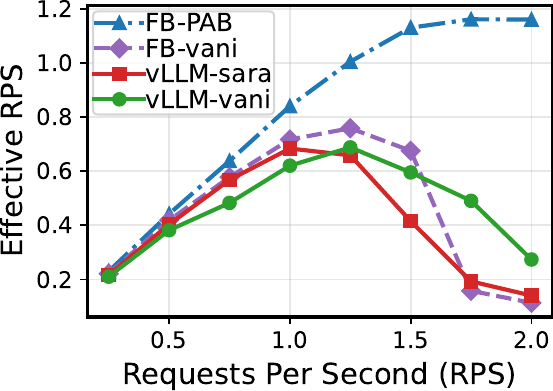}
        \label{fig:subfig_f}
    }
    \hfill
    \subfloat[Llama-3.3-70B; AzureTrace]{
        \includegraphics[width=0.24\linewidth]{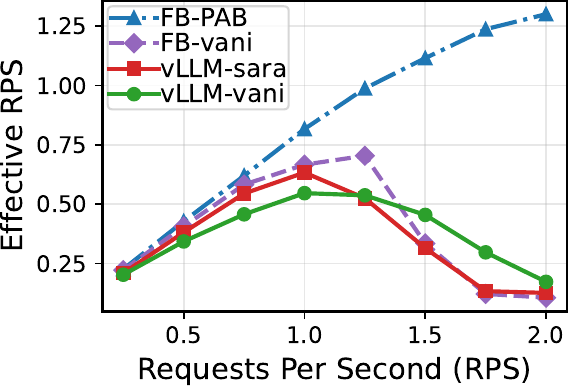}
        \label{fig:subfig_d}
    }

    \caption{Effective RPS of {\sys} under different loads, tested under different models and datasets}
    \label{fig:all_subfigs} 
\end{figure*}

\textbf{Evaluation metrics.}
We evaluate the performance of the tested systems using the following metrics:
\begin{itemize}
    \item \underline{\emph{TTFT}}: time to the first token generation.
    \item \underline{\emph{TPOT}}: average time per output token, we use the max TPOT value among all output tokens of the request to show its worst-case token generation rate.
    \item \underline{\emph{SLO Violation Rate}}: percentage of requests that violate the SLO. In single-node evaluation for PAB, we consider a request to be violated if it is rejected by the PAB, thereby ensuring the fairness of the comparison.
    \item \underline{\emph{Effective RPS}}: effective requests per second, calculated as the product of requests per second and the percentage of requests that satisfy the SLO.
\end{itemize}

\subsection{Single-Node Performance}

In this experiment, we extensively evaluated the performance of four representative systems under varying loads using multiple datasets and model configurations. 
Since our goal is to highlight whether different requests receive fair service under combined TTFT and TPOT objectives, it is insufficient to visualize TTFT and TPOT compliance separately. 
Even if aggregate metrics appear balanced, individual requests may still experience extreme combinations of TTFT and TPOT outcomes, revealing underlying fairness issues.

To address this, we directly measure the number of requests that simultaneously meet both TTFT and TPOT SLOs, and normalize this value into a goodput (effective Requests Per Second) metric. 
This allows us to compare performance across systems based on their ability to satisfy end-to-end SLO requirements.

As shown in Figure~\ref{fig:all_subfigs}, both {\sys} variants achieve higher effective RPS than the two baseline systems across most load levels. 
{\sys}-vanilla achieves higher peak goodput than baselines in all testcases, but only performance degrade under extremely heavy load.
{\sys}-PAB, equipped with admission control to avoid system overload, achieves the highest peak goodput in all test cases under all load levels. 

For quantitative comparison, we extract the peak goodput for each system in every sub-figure and compute the geometric mean across the four models per trace (Table~\ref{tab:peak_goodput}). 
Overall, {\sys}-Vanilla and {\sys}-PAB improve peak goodput over the best baseline(vLLM-sarathi) by 20.0\% and 90.1\%, respectively. The improvement is most pronounced on the Qwen dataset (23.8\% and 109\% over vLLM-vanilla).

\begin{table}[t]
    \caption{Peak goodput achieved by each system, each value represents the geometric mean goodput of 4 evaluated models}
    \centering
    \begin{tabular}{@{}c@{\hspace{5pt}} c @{\hspace{5pt}} c @{\hspace{5pt}} c @{\hspace{5pt}} c@{}}
        \hline
        \multirow{2}{*}{Dataset} & \multicolumn{4}{c}{Peak Effective RPS (Geometric Mean)} \\
        \cline{2-5}
        & vllm-vani & vllm-sara & FB.-vani & FB.-PAB \\
        \hline
        BurstGPT & 1.55 & 1.55 & 1.81 & 2.71 \\
        QwenTrace & 1.22 & 1.13 & 1.51 & 2.56 \\
        AzureTrace & 0.70 & 0.77 & 0.84 & 1.36 \\
        \hline
        Overall & 1.09 & 1.10 & 1.32 & 2.11 \\
        \hline
    \end{tabular}
    \label{tab:peak_goodput}
\end{table}

\subsection{Latency Detail}
 
To illustrate how {\sys} achieves higher system goodput, we compare the latency statistics of {\sys}-vanilla and {\sys}-PAB with the baseline systems on the Qwen3-14B model under the QwenTrace with RPS=2.5.
As Table~\ref{tab:latency_detail} shows, {\sys}-vanilla achieves similar TTFT performance as vLLM-vanilla, while providing stronge TPOT guarantees as vLLM-sarathi. Its P99 TTFT is 2.29$\times$ smaller than vLLM-sarathi, and P99 TPOT is 12.28$\times$ smaller than vLLM-sarathi.

\begin{table}[t]
    \centering
    \caption{Latency statistics of different systems, tested on the Qwen3-14B model under the QwenTrace with RPS=2.5}
    \label{tab:inference_stats}
    \begin{tabular}{lccc ccc}
    \toprule
    \multirow{2}{*}{} & \multicolumn{3}{c}{TTFT (ms)} & \multicolumn{3}{c}{TPOT (ms)} \\
    \cmidrule(lr){2-4} \cmidrule(lr){5-7}
     & P50 & P95 & P99 & P50 & P95 & P99 \\
    \midrule
    vLLM-vani & 212 & 1061 & 1842 & 26 & 256 & 614 \\
    vLLM-sara & 377 & 2455 & 4585 & 29 & 46 & 49 \\
    FB-vani & 213 & 1053 & 1998 & 27 & 50 & 50 \\
    FB-PAB & 198 & 500 & 504 & 23 & 50 & 50 \\
    \bottomrule
    \end{tabular}
    \label{tab:latency_detail}
\end{table}

\begin{figure}[t]
    \centering
    \includegraphics[width=1\columnwidth]{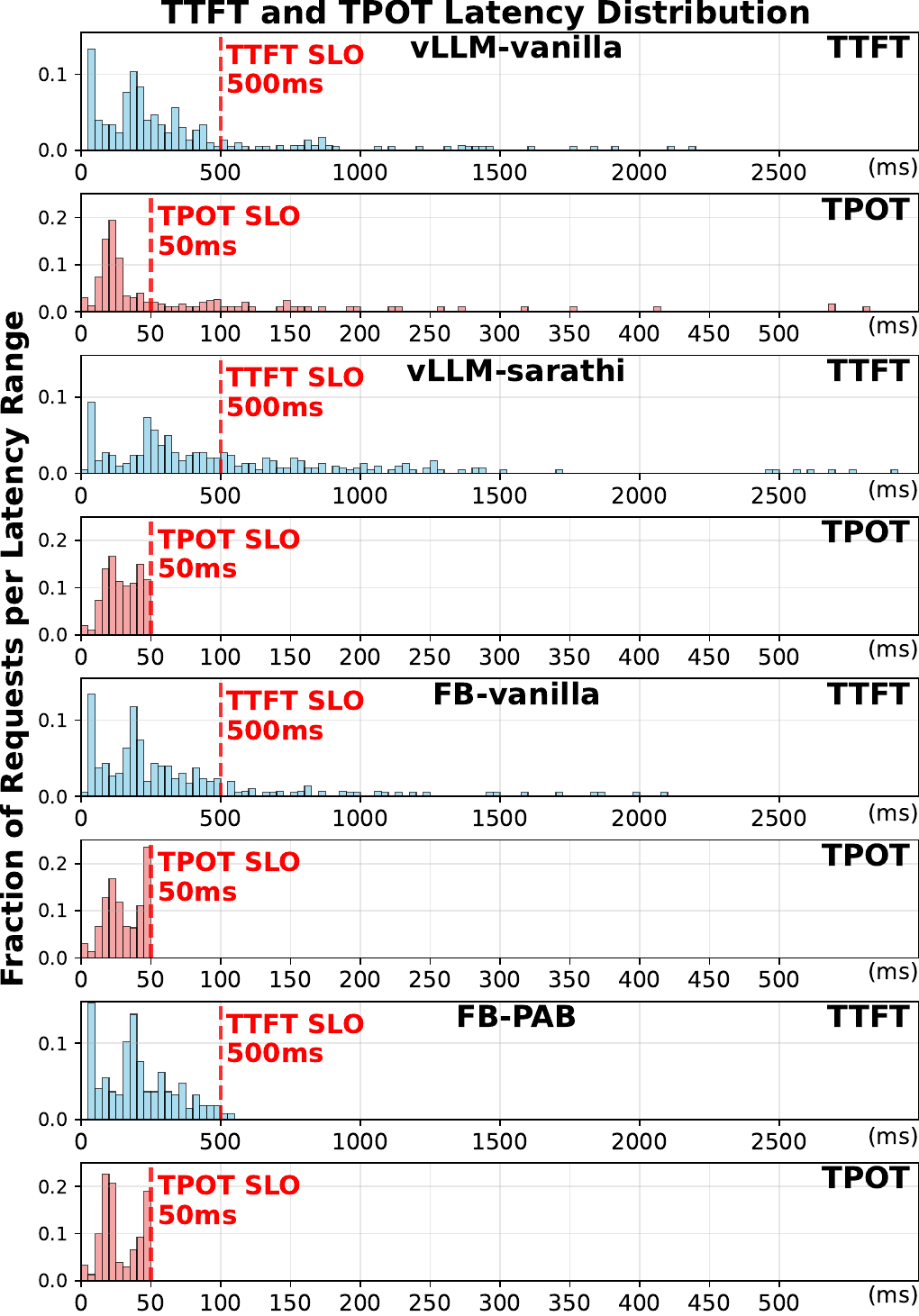}
    \caption{TTFT and TPOT latency distributions of different systems, tested on the Qwen3-14B model under the QwenTrace}
    \label{fig:slo_dist}
\end{figure}

Figure~\ref{fig:slo_dist} further presents a set of frequency distribution plots of request TTFT and TPOT latencies in this experiment.
The results show that while the vLLM-vanilla delivers strong TTFT performance, its aggressive insertion of prefill tasks frequently interrupts decode tasks, leading to probabilistic long pauses in decoding. 
In contrast, the vLLM-sarathi provides strong TPOT guarantees by constraining the maximum batch size per step. 
However, due to its inability to detect excessive decode rate, many requests finish decoding earlier than desired, sacrificing TTFT latency in the process.

{\sys}-vanilla, on the other hand, dynamically constructs each step by monitoring per-request SLO attainment, achieving a TTFT distribution comparable to vLLM-vanilla while providing strong TPOT assurance as vLLM-sarathi. 
It optimizes resource allocation by reducing the compute resources allocated for decode tasks with excessive slack, 
allowing them to run just within SLO requirements if queuing prefill tasks exists. 
This approach frees up capacity and makes {\sys} more robust to prefill bursts.

The remaining SLO violations in {\sys}-vanilla are primarily caused by short-term surges in prefill arrivals. 
To solve this problem, {\sys}-PAB proactively detects when prefill processing capacity is nearing exhaustion and rejects new requests, directing the global scheduler to route them to other nodes. 
This prevents prolonged queuing and ensures that admitted requests are almost guaranteed to meet their SLOs, 
resulting in near-ideal SLO compliance across the system.

\subsection{SLO Versatility}
While the preceding evaluations were conducted under common SLO settings, it is common that different application scenarios may impose varying SLO requirements on LLM inference. 
To examine whether {\sys}'s performance gains persist across different SLO configurations, we conducted an additional experiment where we varied the SLO targets and measured the corresponding improvement over the baseline.

For simplicity, this experiment used the QwenTrace with the Qwen3-14B model. 
The results, shown in Table~\ref{tab:envelope-pab-slo-versatility}, indicate that under the most stringent SLO settings, {\sys}-Vanilla achieves a 15.4\% improvement in peak goodput. 
This performance advantage remains consistent until the SLOs are relaxed to TTFT=2s and TPOT=200ms (equivalent to an output rate of only 5 tokens per second). 
At this point, {\sys}-Vanilla's performance becomes nearly indistinguishable from the baseline.

Analysis of the execution details reveals that the relaxed SLOs allow the baseline system to meet SLO requirements with high probability even under its native scheduling strategy, diminishing the relative advantage of {\sys}-Vanilla. 
However, as shown in Table~\ref{tab:envelope-pab-slo-versatility}, the {\sys}-PAB variant with admission control remains effective even under these lenient conditions. 
By preventing system overload, it ensures a high SLO compliance rate for all admitted requests, demonstrating its effectiveness across a wide range of SLO configurations.

\begin{table}[t]
    \centering
    \caption{{\sys}-vanilla \& {\sys}-PAB's peak goodput improvement over baseline (max of vLLM-sarathi and vLLM-vanilla) under different TTFT and TPOT SLOs}
    \begin{tabular}{ccccc}
    \toprule
    \multicolumn{5}{c}{{\sys}-vanilla Goodput Improvement} \\
    \midrule
        & \multicolumn{4}{c}{TPOT SLO} \\
        \cmidrule(l){2-5}
    TTFT SLO & 50ms & 100ms & 150ms & 200ms \\
    \midrule
    500ms & +15.4\% &  +7.5\% & +5.6\% & +6.1\% \\
    1000ms & +11.6\% &  +8.1\% & +3.9\% &  +2.8\% \\
    1500ms & +8.3\% &  +4.8\% & +2.6\% & +0.6\% \\
    2000ms & +6.8\% &  +4.5\% & +1.5\% & -0.4\% \\
    \bottomrule
    \toprule
    \multicolumn{5}{c}{{\sys}-PAB Goodput Improvement} \\
    \midrule
        & \multicolumn{4}{c}{TPOT SLO} \\
        \cmidrule(l){2-5}
    TTFT SLO & 50ms & 100ms & 150ms & 200ms \\
    \midrule
    500ms & +112.6\% & +91.8\% & +84.9\% & +78.3\% \\
    1000ms & +76.5\% &  +64.9\% & +54.7\% & +55.5\% \\
    1500ms & +63.7\% & +52.1\% & +36.2\% & +32.5\% \\
    2000ms & +59.9\% & +45.9\% & +26.5\% & +24.8\% \\
    \bottomrule
    \end{tabular}
    \label{tab:envelope-pab-slo-versatility}
\end{table}

\subsubsection{Performance Breakdown}
\label{sec:eval-breakdown}

\begin{figure}[t]
    \centering
    \includegraphics[width=1\columnwidth]{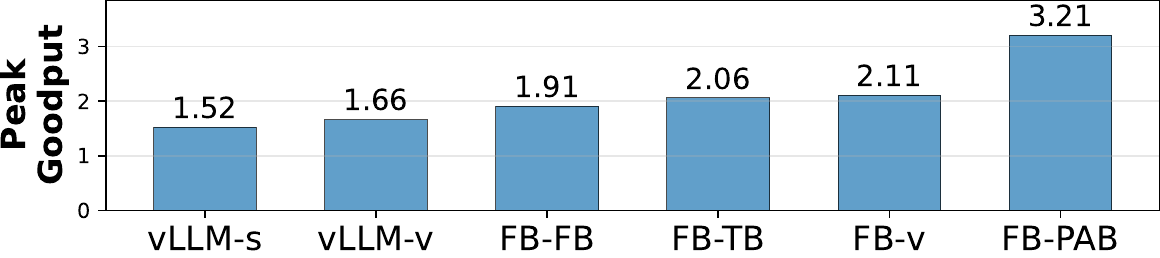}
    \caption{Performance breakdown of different systems}
    \label{fig:breakdown}
\end{figure}

To demonstrate the contribution of each design component, we conducted a performance breakdown analysis using the QwenTrace and Qwen3-14B model, with results illustrated in Figure~\ref{fig:breakdown}. The tested system configurations include: (1) vLLM-sarathi(vLLM-s), (2) vLLM-vanilla(vLLM-v), (3) {\sys}-Fix-Batch(FB-FB, a {\sys} variant with fixed batch size just like vLLM-sarathi, only the token scheduling in Section~\ref{subsec:fair-and-dynamic-token-scheduling} is functional), (4) {\sys}-Token-Budget (FB-TB, a {\sys} variant with a dynamic batch size determined by a dynamic token-based budget system), (5) {\sys}-vanilla (FB-v, similar to {\sys}-vani in previous experiments, change FB-TB's budget to a time-based budget in Section~\ref{subsec:batch-capacity-determination}), and (6) {\sys}-PAB(FB-PAB, a {\sys} variant with additional PAB-based admission control from Section~\ref{subsec:upper-level-scheduler-friendly}).

Results in Figure~\ref{fig:breakdown} show that: (1) vLLM-vanilla achieves 9.2\% higher goodput over vLLM-sarathi, since sarathi's TTFT violation covers its gain in TPOT SLO. (2) FB-FB achieves 15.1\% higher goodput over vLLM-vanilla, since FB-FB allows higher prefill throughput when decodes have slack and provides TPOT SLO guarantees. (3) FB-TB achieves 7.9\% higher goodput over FB-FB, since it can better leverage larger batches when decode slack is abundant, which improves the efficiency of prefill tasks. (4) FB-vanilla achieves 2.4\% higher goodput over FB-TB, since the token-based budget fails to estimate step time accurately when the average context length exceeds expectations, leading to TPOT violations, and FB-vanilla's time-based budget considers the context length and improves the accuracy of step time estimation. (5) FB-PAB further achieves 52.1\% higher goodput over FB-vanilla, since it can better prevent system overload. To this end, FB-PAB achieves 93.4\% higher goodput over the strongest baseline.

\subsection{Cluster-Level Performance}
The final question we need to answer is whether {\sys}'s instance-level scheduler functions effectively when integrated into a distributed inference cluster, and whether the proposed PAB-based load balancer operates as intended. 
To answer this, we designed a new experiment using a data-parallel (DP) group consisting of multiple single-GPU inference engines, each running Qwen3-14B, and evaluated the system across all three datasets.

We compared four system configurations in this experiment:

\begin{itemize}
\item \underline{\emph{vLLM-LB + vLLM-Vanilla}}: The baseline using vLLM's native load balancer (which distributes requests based on active request count) with vLLM's default scheduler.

\item \underline{\emph{vLLM-LB + vLLM-Sarathi}}: The baseline combining the same vLLM load balancer with the Sarathi scheduler.

\item \underline{\emph{vLLM-LB + {\sys}-Vanilla}}: A hybrid configuration using vLLM's load balancer but replacing the node-level scheduler with {\sys}-Vanilla. This tests {\sys}'s distributed performance without PAB-aware load balancing.

\item \underline{\emph{PAB-LB + {\sys}}}: The full {\sys} system, integrating the PAB-based load balancer with {\sys} node schedulers.
\end{itemize}

\begin{figure}[h]
    \centering
    \subfloat[BurstGPT Data-Parallel peak Goodput]{
        \includegraphics[width=\linewidth]{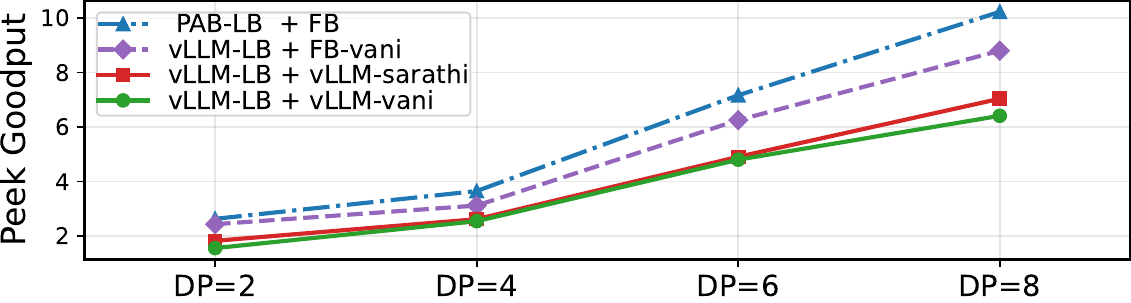}
    }\\
    \subfloat[QwenTrace Data-Parallel peak Goodput]{
        \includegraphics[width=\linewidth]{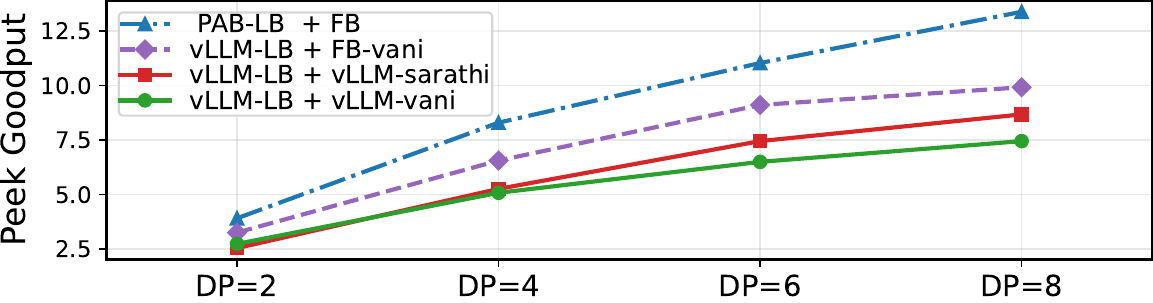}
    }\\
    \subfloat[AzureTrace Data-Parallel peak Goodput]{
        \includegraphics[width=\linewidth]{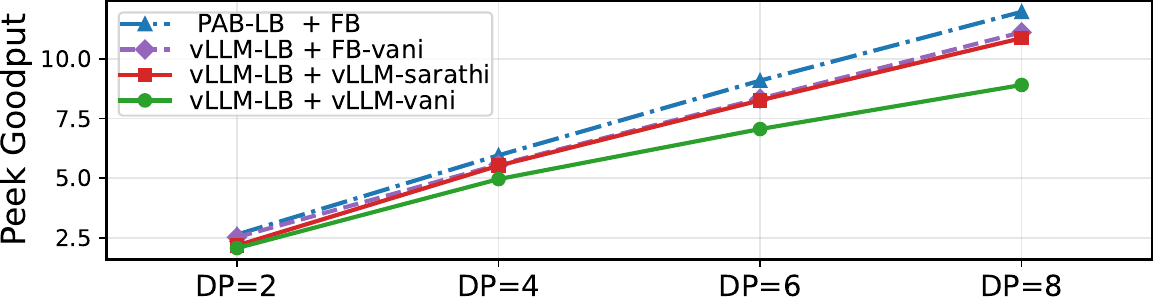}
    }
    \caption{peak goodput of different systems and different load balancing strategies.}
    \label{fig:all_subfigs_dp}

\end{figure}

Results in Figure~\ref{fig:all_subfigs_dp} demonstrate that the PAB-LB provides an additional 34.9\%, 16.2\%, and 7.7\% (QwenTrace, BurstGPT, AzureTrace) performance improvement to the {\sys} system at DP=8, leading to an overall gain of 54.3\%, 45.3\%, 38.3\%, and 10.2\% (QwenTrace, BurstGPT, AzureTrace) compared to the best baseline.

\section{Related work}
\label{sec:related}
\subsection{SLO Aware LLM Serving Systems}

Beyond the detailed discussion of Sarathi-Serve~\cite{sarathi} in the main text, a number of other works have explored scheduling for LLM inference with the goal of maximizing SLO attainment.
VTC~\cite{sheng2024fairness} proposed a token-based system for fair request admission, while QLM~\cite{patke2024queue} and Prism~\cite{yu2025prism} optimized multi-model serving through solver-based scheduling and memory-aware extensions.
LightLLM~\cite{gong2025past} and SELECT-N~\cite{ma2025memory} addressed memory-related SLO violations through length prediction and opportunistic swapping.
ExeGPT~\cite{oh2024exegpt} and SCOOT~\cite{cheng2024towards} automated inference engine configuration search, complemented by auto-scaling approaches like Chiron~\cite{patke2025hierarchicalautoscalinglargelanguage}.
Energy-efficient scheduling under SLO constraints has been explored in~\cite{liang2023energy, qiu2024power, 10540202, wilkins2024hybrid, mammen2023cuff, liang2023dvelen}, while performance prediction~\cite{258862} and fairness-aware methods~\cite{cao2025localityawarefairschedulingllm} further contribute to SLO attainment.

While these works have made significant contributions to SLO-aware LLM inference, a common limitation is their treatment of the core inference engine and its step-level scheduler as a black box. They often overlook the fine-grained, per-step, and per-request performance characteristics intrinsic to the decoding process. {\sys} offers a novel perspective by directly addressing the inter-request unfairness that arises within a single instance, unlocking a new dimension for performance gains that operates beneath the abstraction level of most prior systems.

\subsection{Prefill-Decode Disaggregation Systems}
Several other works adopt prefill-decode (PD) disaggregation to meet Service Level Objectives (SLOs) for LLM serving. DistServe~\cite{distserve} optimizes serving by disaggregating prefill and decode with phase-specific resource allocation. Splitwise~\cite{patel2024splitwise} separates prompt and generation across machines for better utilization. TetriInfer~\cite{hu2024inference} handles mixed workloads via disaggregated inference, while MemServe~\cite{hu2024memserve} manages memory and KV caches with an elastic pool to reduce JCT and TTFT. Mooncake~\cite{mooncake} introduces a KV-cache-centric design with SLO-aware scheduling, and LoongServe~\cite{mooncake} leverages Elastic Sequence Parallelism to adapt parallelism across phases.
Recent innovations in this space include TaiChi~\cite{wang2025prefill}, which dynamically switches between PD-disaggregated and PD-collocated modes for goodput improvement.
In contrast, {\sys} adopts a PD-collocated architecture, which avoids KV-cache transfer overhead, enables larger batches when slack exists, and removes the need for cross-cluster adjustment under shifting load ratios~\cite{feng2025windserve}, while still achieving comparable two-stage latency guarantees.
\subsection{LLM Serving Optimizations}
Research on LLM serving optimizations includes 
Orca~\cite{orca}'s continuous batching which allows new requests joining ongoing batches and improves overall throughput. 
vLLM~\cite{pagedattention} introduces PagedAttention which dramatically reduces KV cache fragmentation and enables near-optimal memory utilization. NanoFlow~\cite{zhu2025nanoflow} implements nano-batching and operation overlapping to further improve inference throughput. Additional advances include attention disaggregation~\cite{chen2025efficientheterogeneouslargelanguage, hamadanian2025glinthawk, liang2025injectingadrenalinellmserving}, speculative decoding~\cite{cho2025lossless}, memory offloading strategies~\cite{jiang2024neosavinggpumemory}, and adaptive quantization~\cite{su2025efficient}. In the domain of system architecture, CloudMatrix-Infer~\cite{zuo2025servinglargelanguagemodels} employs a hyper-node architecture for efficient serving. Furthermore, serverless deployments are gaining traction, as seen in systems like ServerlessLLM~\cite{298681}, BlitzScale~\cite{zhang2025blitzscale} and Medusa~\cite{zeng2025medusa}, the last of which utilizes CUDA Graph materialization techniques to accelerate serverless LLM.

In this landscape of increasingly refined optimizations, {\sys} introduces a scheduling mechanism compatible with these techniques. By ensuring fair compute allocation among requests, it achieves higher fairness and maximizes the utility of existing LLM serving systems.
\section{Conclusion}
\label{sec:concl}

This paper introduces FairBatching, a scheduling system that ensures fair quality-of-service in LLM inference. 
By dynamically allocating resources based on fine-grained per-request SLO tracking, FairBatching significantly reduces tail latency and improves goodput in both single-node and distributed settings, demonstrating that explicit fairness control is key to unlocking higher efficiency in modern serving systems.

\balance

{\footnotesize 
\bibliographystyle{ACM-Reference-Format}
\bibliography{main}}

\newpage

\appendix
\section{Appendix: Prefill Admission Budget (PAB) Calculation}
\label{appendix:pab}

This appendix provides a step-by-step derivation of the Prefill Admission Budget (PAB) formula, which estimates the number of additional prefill tokens a node can process within a new request's Time-To-First-Token (TTFT) Service Level Objective (SLO).

\subsection{Step 1: Total Computational Time within TTFT}
The total available computational time within the TTFT window is represented by the time budget $TTFT$ itself.

\subsection{Step 2: Resource Reserved for the Most Urgent Decode Task}
In the worst case, the decode task with the smallest slack ($\min_{i \in Decode} slack_i$) must be scheduled within $TTFT_{slo} - \min_{i \in Decode} slack_i$ time units to avoid SLO violation. Then, for each TPOT window, this urgent decode task will take at least one step. So the total number of steps within the TTFT window is:
\[
N_{batches} =  \frac{TTFT_{slo} - \min_{i \in Decode \cup Prefill} slack_i}{TPOT_{slo}}  + 1
\]
Each step incurs a fixed overhead of $a$. Thus, the total time reserved for the most urgent decode task is:
\[
R_{batches} = N_{batches} \cdot a
\]

\subsection{Step 3: Resource Consumption by decode}
Each request $i$ (both decode and prefill) must complete a certain number of decode steps within $TTFT_{slo}$ if $TTFT_{slo} > slack_i$. The number of decode steps for request $i$ is:
\[
N_i =  \frac{TTFT_{slo} - slack_i}{TPOT_{slo}}
\]
The compute time per step for request $i$ depends on its context length: $b + context_i \cdot c$. The total resource consumed by all decode tasks is (here we ignore the context increment to simplify the calculation):
\[
R_{tasks} = \sum_{i \in Decode \cup Prefill} N_i \cdot (b + context_i \cdot c)
\]

\subsection{Step 4: Remaining Resource for New Prefill Task}
The computational resource remaining for prefill tasks is:
\[
R_{prefill} = TTFT_{slo} - R_{urgent} - R_{tasks}
\]

\subsection{Step 5: Token Capacity of the Remaining Resource}
The new prefill task's processing time per token is modeled as $b + c$ (the context length is initially just the prompt tokens, so is the same as the prompt tokens, here we ignore the additional attention overhead if a prefill is segmented to multiple chunks, this is a reasonable approximation since for prefill tasks, the token-related performance factor is dominant). The number of new prefill tokens that can be processed with $R_{prefill}$ time is:
\[
T_{prefill} = \frac{R_{prefill}}{b + c}
\]

\subsection{Step 6: Adjusting for Existing Prefill Tokens}
$T_{new}$ represents the total prefill token capacity. To find the \textit{admission budget}, we subtract the number of tokens from existing, unfinished prefill tasks:
\[
PAB = T_{prefill} - \sum_{i \in Prefill} prompt_i
\]

\subsection{Final Formula}
Combining all steps and approximating yields the PAB formula:
\[
\begin{aligned}
PAB = & \frac{1}{b + c} \left[ TTFT_{slo} - \left[\frac{ (TTFT_{slo} - \min\limits_{i \in D \cup P} slack_i)}{TPOT_{slo}}  + 1 \right]  \cdot a \right. \\
& \left. - \sum_{i \in D \cup P} \frac{TTFT_{slo} - slack_i}{TPOT_{slo}} \cdot (b + context_i \cdot c) \right] \\
& - \sum_{i \in Prefill} prompt_i
\end{aligned}
\]
This formula allows the upper-level scheduler to make informed routing decisions based on a locally maintained, approximate view of each node's capacity.

\end{document}